\documentstyle[eqsecnum,aps,prd]{revtex}

\begin{document}




\thispagestyle{empty}
{\baselineskip0pt
\leftline{\large\baselineskip16pt\sl\vbox to0pt{\hbox{Department of Physics} 
               \hbox{The University of Tokyo}\vss}}
\rightline{\large\baselineskip16pt\rm\vbox to20pt{\hbox{UTAP-252}
               \hbox{RESCEU-12/97}
               \hbox{\today}
\vss}}%
}
\vskip15mm

\begin{center}
{\large\bf  Freely Falling 2-Surfaces and the Quasi-Local Energy}
\end{center}

\begin{center}
{\large Keita Ikumi} \\
\sl{Department of Physics, The University of Tokyo, Tokyo 113, Japan}
\end{center}
\begin{center}
{\large Tetsuya Shiromizu\footnote
{Present Address:DAMTP, Univ. of Cambridge, 
Silver Street, Cambridge 3DN 9EW, UK}} \\
\sl{Department of Physics, The University of Tokyo, Tokyo 113, Japan \\
and \\
Research Center for the Early Universe(RESCEU), \\ The University of Tokyo, 
Tokyo 113, Japan
}
\end{center}

\begin{abstract}
We derive an expression for effective gravitational mass for any
closed spacelike 2-surface. This effective gravitational energy is
defined directly through the geometrical quantity of the freely
falling 2-surface and thus is well adapted to intuitive expectation
that the gravitational mass should be determined by the motion of test
body moving freely in gravitational field. We find that this effective
gravitational mass has reasonable positive value for a small sphere 
in the non-vacuum space-times and can be negative for vacuum case. 
Further, this effective
gravitational energy is compared with the quasi-local energy based on
the $(2+2)$ formalism of the General Relativity. Although some gauge
freedoms exist, analytic expressions of the quasi-local energy for
vacuum cases are same as the effective gravitational mass. 
Especially, we see that the contribution from the cosmological
constant is the same in general cases.
\end{abstract}

\section{Introduction\label{intro}}

One frequently wants to define the local energy in order to
investigate the local structure of the dynamical spacetimes. However,
it is well known that, due to the equivalence principle, the
gravitational field does not have local (point-wise) energy density in
General Relativity. Hence, it might be impossible to construct the
combined energy density of gravity and matter in a purely local
manner.

Fortunately, for asymptotically flat spacetimes, it was shown that the
notion of the total energy for the whole 3-space exists and that one
can consistently define the total energy at spatial and null infinity:
the ADM energy $E_{\rm ADM}$\cite{ADM} and the Bondi-Sachs energy
$E_{\rm BS}$\cite{Bondi,Sachs}, respectively. They have several nice
features. They are defined in entirely covariant
ways\cite{AH}. The positivity of these energy related to the stability
of spacetimes has been proven by \cite{SY,Witten}. The relation
between $E_{\rm ADM}$ and $E_{\rm BS}$ is revealed in Ref.\cite{AMA}

In the case of asymptotically de Sitter spacetimes, one can define the
Abott-Deser energy\cite{AD}. This is an integral of the conserved
charge and has the nature such as the total energy of the whole
3-space. In spherical cases, Nakao, Shiromizu and Maeda\cite{NSM}
showed that the Abott-Deser energy picks up correctly the
gravitational mass, which determines the tidal force. Further, they
found an example in which the Abott-Deser energy is negative. One of
the present authors showed that it is positive in cases corresponding
with `static' like spacetimes\cite{TS}.

The above success of these total energies in spacetimes with different
asymptotic structure urges again people to construct the local notion
of the gravitational energy, that is, one want to define the useful
total energy of compact regions which does not depend on the
asymptotic structure of spacetimes\cite{QLM}.

In the present paper, we consider the motion of a freely falling
2-surface and then define the effective gravitational mass for that
surface. Then we evaluate it for small spheres in non-vacuum
space-times and in Schwarzschild space-time. Furthermore, we show that it
resembles the quasi-local energy derived from the total Hamiltonian of
the $(2+2)$ formalism.

The organization of the present paper is as follows. In Section 2, we
introduce the effective gravitational mass for a freely falling
2-surface in General Relativity based on the comparison with the
Newtonian theory and evaluate it for small spheres. We also give
explicit values for some known spacetimes, most notable of which are a
small sphere outside the black hole and large sphere containing the
black hole in the Schwarzschild spacetime. In Section 3, we show that
our effective gravitational energy for a 2-surface has similarity with
the quasi-local energy derived from the $(2+2)$ formalism, first
reviewing that formalism briefly. Finally, we give the summary in
Section 4.

\section{Gravitational mass and freely falling 2-surface\label{interp}}

In the Newtonian theory of gravity, the motion of a test particle with
vanishing angular momentum is determined by
%
\begin{equation}
{\ddot r} = -\frac{M}{r^2}.
\label{Newtoneq}
\end{equation}
%
This equation relates the gravitational mass $M$ with the
time evolution of a geometrical quantity ${\ddot r}$. We seek for a similar
relation in the context of General Relativity. 
In the following, we first establish the mass-geometry relation
in terms of the quantities associated with freely falling 2-surface in
Newtonian theory and then extend that relation to the case of General
Relativity.

\subsection{Effective mass in Newtonian theory\label{effective}}

We consider an arbitrary 2-surface $S$ in the Newtonian absolute
spacetime. Imagine that $S$ is entirely covered with freely infalling
test particles. Denote the tangent vector of the world line of each
test particle by $t^a$ and define the 3-velocity $v^a$ by $t^a =
(\partial / \partial t)^a + v^a$, where $t$ is the absolute time. 
Write the flat metric of the absolute space as $g_{ab}$ and the
natural derivative operator of the spacetime as $\nabla_a$. Then the
equation of motion takes the form
%
\begin{displaymath}
t^a \nabla_a v^b = - g^{ab} \nabla_a \Phi,
\end{displaymath}
%
where $\Phi$ is the gravitational potential which satisfies the 
Poisson equation
%
\begin{equation}
g^{ab}\nabla_a \nabla_b \Phi = 4\pi \rho,
\label{Poissoneq}
\end{equation}
%
where $\rho$ is the matter density.

Denote the projection tensor onto $S$ by $h_a^b$, where $h_{ab} \equiv
h_a^c g_{cb}$ is the metric of $S$. These metrics satisfy $g^{ab} =
h^{ab} + n^a n^b$ where $n^a$ is the normal vector of $S$ pointing
outwardly and $h^{ab} \equiv g^{ac}h_c^b$. Define ${\hat \theta}$ by $h_a^c
h_b^d \pounds_t \mu_{cd} = {\hat \theta} \mu_{ab}$ where $\mu_{ab}$ is
the area 2-form of $S$. Then ${\hat \theta} = h_b^a \nabla_a t^b =
h_b^a \nabla_a v^b$ gives the evolution rate of the infinitesimal area
element $\delta A$ which consists of some fixed members of the test
particles: ${\hat \theta} = t^a \nabla_a (\delta A)/\delta A$. 
${\hat \theta}$ is also the expansion of the congruence of
the test particles. Next, introduce the shear ${\hat \sigma}_{ab}
\equiv 2h_a^c h_b^d\nabla_{(c} v_{d)} - {\hat \theta} h_{ab}$ and the
rotation ${\hat \omega}_{ab} \equiv 2h_a^c h_b^d \nabla_{[c} v_{d]}$
of the congruence. Then we can
determine time evolution of $t^b\nabla_bn_a$ as follows.

Since $t^b\nabla_bn_a$ gives zero when contracted with
$(\partial/\partial t)^a$ and
$n^a$, it has only components parallel to $S$. Take an arbitrary
vector field $X^a$ tangent $S$. Then we have
\begin{eqnarray}
X^at^b\nabla_bn_a &=& -(t^b\nabla_bX^a)n_a \nonumber \\
&=& -(\pounds_tX^a + X^b\nabla_bt^a)n_a \nonumber\\
&=& -X^bn_a\nabla_bv^a.
\label{evolutionofn}
\end{eqnarray}
The last equality holds because $S$ is Lie-propagated along $t^a$,
which is assumed implicitly in the above construction. Since $X^a$ is
arbitrary other than $X^a$ should be tangent to $S$, we have the
relation $t^b \nabla_b n_a = - n_b h_a^c\nabla_c v^b$. Now we can
obtain the time evolution of ${\hat \theta}$ as follows.
%
\begin{eqnarray}
t^c \nabla_c {\hat \theta} &=& t^c\nabla_c(h_b^a\nabla_av^b) \nonumber \\
	&=& (t^c\nabla_ch_b^a)\nabla_av^b + h_b^at^c\nabla_c\nabla_av^b\nonumber  \\
	&=& t^c\nabla_c(g_b^a - n_bn^a)\nabla_av^b +
		h_b^a\Bigr(\nabla_a(t^c\nabla_cv^b) - (\nabla_at^c)
                   (\nabla_cv^b)\Bigl)\nonumber \\
	&=& n_d(h_b^c\nabla_cv^d)n^a\nabla_av^b +
		n_bn_d(h^{ac}\nabla_cv^d)\nabla_av^b \nonumber \\
	& & {}+ h_b^a(-\nabla_a(g^{bc}\nabla_c\Phi) -
		(\nabla_av^c)(\nabla_cv^b)) \nonumber\\
	&=& - h_b^ah_d^c(\nabla_av^d)(\nabla_cv^b) +
		h^{ab}(n_c\nabla_av^c)(n_d\nabla_bv^d)\nonumber  \\
	& & {}- h_b^a\nabla_a(h^{bc}\nabla_c\Phi + n^bn^c\nabla_c\Phi)\nonumber  \\
	&=& -\frac14({\hat \theta}h_a^b + {\hat \sigma}_a^b + {\hat \omega}_a^b
)
		({\hat \theta}h_b^a + {\hat \sigma}_b^a + {\hat \omega}_b^a) \nonumber \\
	& & {}+ h^{ab}(n_c\nabla_av^c)(n_d\nabla_bv^d) -
		D_a(h^{ab}D_b\Phi) - h_b^a(\nabla_an^b)n^c\nabla_c\Phi. 
\end{eqnarray}
Thus we have an equation
\begin{equation}
t^a \nabla_a {\hat \theta} + \frac12 {\hat \theta}^2 + \frac14
{\hat \sigma}_{ab} {\hat \sigma}^{ab} - \frac14 {\hat \omega}_{ab}
{\hat \omega}^{ab} = - {\underline \theta} n^a \nabla_a \Phi - D_a
(h^{ab} D_b \Phi) + h^{ab} (n_c \nabla_a v^c) (n_d \nabla_b v^d),
\label{Newtonkey}
\end{equation}
%
where ${\underline \theta} \equiv h^{ab}\nabla_a n_b$ is the trace of
the extrinsic curvature of $S$ in the Euclid space. 
This equation resembles the
Raychaudhuri's equation. Integrating the above equation over the
closed 2-surface $S$, we have
%
\begin{equation}
\int_S \mu \Bigl( t^a \nabla_a {\hat \theta} + \frac12 {\hat \theta}^2
+ \frac14 {\hat \sigma}_{ab} {\hat \sigma}^{ab} - \frac14 {\hat
\omega}_{ab} {\hat \omega}^{ab} \Bigr) = \int_S \mu
\Bigl[ - {\underline \theta} n^a \nabla_a \Phi + h^{ab} (n_c \nabla_a v^c)
(n_d \nabla_b v^d) \Bigr].
\label{basicNewton}
\end{equation}
%
The first term on the right hand side basically gives the material
mass inside $S$ and the second term corresponds to the contribution of
centrifugal force. This is most easily seen for the initial
configuration in which $S$ is a sphere of constant radius $r$ and
$v^a$ is tangent to $S$. For such surface, ${\underline \theta} = 2/r$
and $n_c h^{ab} \nabla_b v^c = - v^c h^{ab}
\nabla_b n_c = - v^a / r$. Thus the value of the right hand side of
(\ref{basicNewton}) is, considering eq.(\ref{Poissoneq}),
%
\[ \int_S \mu \Bigl( - \frac{2}{r} n^a \nabla_a \Phi +
\frac{v^2}{r^2} \Bigr) = 8\pi \Bigl(- \frac{M}{r} + \frac{l^2}{2r^2}\Bigr),\]
%
where $M \equiv \int_V dV \rho$ with $V$ being the region inside $S$
and $l \equiv (\int_S \mu v^2/4\pi)^{1/2}$ is the averaged angular
momentum of the test particles of unit mass. This is identical to (up
to a numerical factor $8\pi$) an effective potential of the test
particles with the angular momentum $l$.

 From these considerations, in general cases, we can define the effective 
mass for general closed surface $S$ by 
%
\begin{eqnarray}
M_{\rm eff}(S) & \equiv & - \Bigl( \frac{A}{4\pi} \Bigl)^{\frac12} \int_S
\frac{\mu}{8\pi} \Bigl( t^a \nabla_a {\hat \theta} + \frac12 {\hat \theta}^2
+ \frac14 {\hat \sigma}_{ab} {\hat \sigma}^{ab} - \frac14 {\hat
\omega}_{ab} {\hat \omega}^{ab} \Bigr). \nonumber \\
               & = &  - \Bigl( \frac{A}{4\pi} \Bigl)^{\frac12} \int_S
\frac{\mu}{8\pi} \Bigl[ - {\underline \theta} n^a \nabla_a \Phi + h^{ab} 
(n_c \nabla_a v^c)(n_d \nabla_b v^d)\Bigr]
\label{effNewton}
\end{eqnarray}
%

\subsection{Effective mass in General Relativity}

Now we consider a freely falling 2-surface in General Relativity. Let
$t^a$ be a unit timelike vector field orthogonal to $S$. Extend each
$t^a$ to be geodesic along its direction. 
As a result we have a
two-dimensional geodesic congruence which starts from $S$. Denote the
metric of the 2-surface which is Lie-propagated along $t^a$ by
$h_{ab}$. Then $t^a$ is always orthogonal to $S$. We define the
expansion of the congruence as ${\hat \theta} \equiv h^{ab} \nabla_a
t_b$. The expansion ${\hat \theta}$ again satisfies $h_a^c h_b^d
\pounds_t \mu_{cd} = {\hat \theta} \mu_{ab}$ and equals to the
evolution rate of the infinitesimal area element which is spanned by
some fixed members of the congruence: ${\hat \theta} = t^a \nabla_a
(\delta A) /\delta A$. In this case the rotation ${\hat \omega}_{ab}
\equiv 2h_a^c h_b^d \nabla_{[c} t_{d]}$ of the congruence vanishes
since $t^a$ is orthogonal to $S$. Thus the shear is given by
${\hat \sigma}_{ab} = 2h_a^c h_b^d \nabla_c t_d - {\hat
\theta}h_{ab}$, without symmetrization over the indices $c$ and $d$.

We can show, by the similar argument to the one in the Newtonian case,
that the evolution of the unit spacelike vector $n^a$ orthogonal to both
$h_{ab}$ and $t^a$ is given by the equation $t^c \nabla_c n^a = - n_b
h^{ac} \nabla_c t^b$.
Then the evolution of ${\hat \theta}$ is given by
%
\begin{eqnarray}
t^c \nabla_c {\hat \theta} &=& t^c\nabla_c(h_b^a\nabla_at^b) \nonumber \\
	&=& (t^c\nabla_ch_b^a)\nabla_at^b + h_b^at^c\nabla_c\nabla_at^b \nonumber \\
	&=& t^c\nabla_c(g_b^a + t_bt^a - n_bn^a)\nabla_at^b +
		h_b^at^c([\nabla_c, \nabla_a]t^b + \nabla_a\nabla_ct^b) \nonumber \\
	&=& n_d(h_b^c\nabla_ct^d)n^a\nabla_at^b +
		n_bn_d(h^{ac}\nabla_ct^d)\nabla_at^b \nonumber \\
	& & {}+ h_b^a\Bigl(t^cR^b_{dca}t^d -
		(\nabla_at^c)(\nabla_ct^b)\Bigr) \nonumber \\
	&=& - h_b^ah_d^c(\nabla_at^d)(\nabla_ct^b) +
		h^{ab}(n_c\nabla_at^c)(n_d\nabla_bt^d) -
		h^{ab}R_{acbd}t^ct^d.\\
\end{eqnarray}
%
The curvature term $h^{ab}R_{acbd}t^ct^d$ is related to the
parallel component of Weyl tensor as
%
\begin{equation}
R_{acbd}h^{ab}t^ct^d = \frac12 C_{acbd}h^{ab}h^{cd} + R_{ab}t^at^b - 
\frac12 R_{ab}h^{ab} +  \frac13R,
\end{equation}
%
so we obtain 
\begin{displaymath}
t^c \nabla_c {\hat \theta} + \frac12 {\hat \theta}^2 +
\frac14 {\hat \sigma}_{ab} {\hat \sigma}^{ab} =
- \frac12 C_{abcd}h^{ac}h^{bd} - R_{ab}t^a t^b + \frac12 R_{ab}h^{ab} -
\frac{R}{3} + \omega^a \omega_a,
\end{displaymath}
where $\omega_a \equiv n_ch_a^b\nabla_bt^c $ and it expresses the
``centrifugal force component.'' This equation should be compared to
(\ref{Newtonkey}). Thus, in the case of General Relativity, from the
Newtonian analogy of eq.(\ref{effNewton}), we can define the effective
mass for the freely falling 2-surface $S$ as
%
\begin{eqnarray}
M_{\rm eff}(S) &=& 
- \Bigl( \frac{A}{4\pi} \Bigl)^{\frac12} \int_S
\frac{\mu}{8\pi} \Bigl( t^a \nabla_a {\hat \theta} + \frac12 {\hat \theta}^2
+ \frac14 {\hat \sigma}_{ab} {\hat \sigma}^{ab} \Bigr) \nonumber\\
&=&
\Bigl( \frac{A}{4\pi} \Bigr)^{\frac12}
\int_S \mu \biggl[
\frac{1}{16\pi} \Bigl( C_{abcd}h^{ac}h^{bd} - 2\omega_a \omega^a -
\frac43 \Lambda \Bigr)
+ T_{ab}t^a t^b - \frac12 T_{ab}h^{ab} + \frac23 T_a^a \biggr],
\label{effectivemass}
\end{eqnarray}
%
where we have used the Einstein equations $R_{ab} - g_{ab}R/2 = 8\pi
T_{ab} - \Lambda g_{ab}$. We propose the effective mass as the quasi-local 
energy. 
We note that the effective mass $M_{\rm eff}(S)$ can be defined for
{\em any\/} 2-surface $S$ although the definition is based on the
freely falling test particles, because, given any 2-surface $S$, the
above argument can be applied to the sequence of 2-surfaces generated
by the motion of freely falling test particles which start off from
that particular 2-surface.

One can show that the effective mass $M_{\rm eff}(S)$ coincides
exactly with the ADM energy and the Bondi-Sachs energy with
appropriate limits in asymptotically flat spacetime. Both of the ADM
energy and the Bondi-Sachs energy can be expressed as an asymptotic
limit of the integral $(\frac{A}{4\pi})^{1/2}
\int_S \frac{\mu}{16\pi} C_{abcd}h^{ac}h^{bd}$.\cite{Hay2,AH,Ash}.
Taking into consideration the asymptotic
behaviour of the twist $\omega_a \omega^a \sim {\cal
O}(r^{-6})$\cite{Hay2}, the standard falloff conditions of the
energy-momentum tensor in the asymptotic region tell us that $E_{\rm
ADM}$ and $E_{\rm BS}$ are given as the appropriate limits of the
effective mass:
%
\begin{displaymath}
E_{\rm ADM} = \lim_{S \to i^0} M_{\rm eff}(S),\quad
E_{\rm BS} = \lim_{S \to {\cal J}} M_{\rm eff}(S).
\end{displaymath}
%

We give the explicit values of $M_{\rm eff}(S)$ for various exact
solutions with spherical symmetry here. In all cases $S$ is taken to
be a sphere of symmetry in $t=\mbox{constant}$ surface:
\begin{eqnarray}
M_{\rm eff}(S) &=& 0 \qquad\mbox{(Minkowski)}\nonumber  \\  
     &=& M \qquad\mbox{(Schwarzschild)}\nonumber  \\  
     &=& M - \frac{e^2}{r} \qquad\mbox{(Reissner-Nordstr\"om)}\nonumber  \\ 
     &=& M - \frac{\Lambda}{3}r^3 \qquad\mbox{(Schwarzschild-de
Sitter)}\nonumber  \\
     &=& \frac{4\pi}{3}r^3 (\rho + 3P) - \frac{\Lambda}{3}r^3
\qquad\mbox{(Friedmann-Robertson-Walker)}.
\end{eqnarray}
All these values satisfy eq.(\ref{Newtoneq}) for radial timelike
geodesic regarding $M_{\rm eff}(S)$ as gravitational mass, as is
expected from the above derivation, since $M_{\rm eff}(S)$ is defined
directly through the behaviour of freely falling 2-surface. Actually,
one can see that $M_{\rm eff}(S)$ always gives the correct
gravitational mass in the above sense in spherically symmetric
spacetimes when $S$ is a symmetric sphere. Proof: In such spacetimes,
the area of $S$ is expressed as $A = 4\pi r^2$ and
the shear ${\hat \sigma}_{ab}$ vanishes. In addition, ${\hat \theta}$
is constant over $S$ and equals to ${\dot A}/A = 2{\dot r}/r$. Thus
the effective energy is easy to compute and gives $M_{\rm
eff}(S)=-r^2{\ddot r}$.

Here we mention two features of $M_{\rm eff}(S)$. First, in the
Reissner-Nordstr\"om (RN) spacetime, it does not coincide with the
Misner-Sharp energy\cite{MS}, which has been widely accepted as the
correct quasi-local energy in the spherically symmetric models so far,
$E_{\rm MS} = M - e^2/2r$. Secondly, $M_{\rm eff}(S)$ does not
coincide with the Abott-Deser energy in the Schwarzschild-de Sitter
(SdS) spacetime and diverges to negative infinity as $r \to \infty$. 
The latter feature is not difficult to understand in the current
context. In the region far away the black hole, the nature of the SdS
spacetime is approximately same as the de Sitter spacetime. Thus the
freely falling sphere staying in that region actually flies away from
the center due to the cosmological rapid expansion. So the effective
mass in this case should be negative. Since the expansion is caused by
uniformly distributed vacuum energy $\Lambda$, it is natural that
$M_{\rm eff}(S)$ has $\Lambda r^3$ dependence. Although these features
seem queer at first glance, $M_{\rm eff}(S)$ surely captures some of
the features of the gravitational mass. From the construction one can
easily see that it is useful to consider the dynamics of the compact
object.

\subsection{The evaluation on small spheres}
 
A few years ago Bergqvist\cite{Small} studied the energy of small
spheres and showed that Hayward's energy\cite{Hay2} becomes negative
for a small sphere in vacuum case. The effective mass $M_{\rm eff}$
agrees with the Hayward energy in vacuum spacetimes as shown in the
next section, so we investigate the properties of $M_{\rm eff}$ for
small spheres in this subsection.

First, we consider non-vacuum case, that is, $T_{ab}\neq 0$. In this 
case, as the same way of Bergqvist's estimation, 
one can estimate the present effective gravitational mass 
easily. The leading term is given by 
\begin{eqnarray}
M_{\rm eff}(S) & \sim & \frac{4\pi}{3}r^3(T_{ab}t^at^b+T_{ab}q^{ab}) 
-\frac{\Lambda}{3}r^3\\ 
\nonumber \\
& = & \frac{r^3}{3}R_{ab}t^at^b,
\end{eqnarray}
where $q_{ab}$ is the metric of the hypersurface orthogonal to 
$t^a$. If one defines the effective local energy density and pressure 
by 
\begin{eqnarray}
\rho_{\rm eff}:
=T_{ab}t^at^b\quad\mbox{and}\quad P_{\rm eff}:=\frac{1}{3}T_{ab}q^{ab},
\end{eqnarray}
the expression of the leading term becomes
\begin{eqnarray}
M_{\rm eff}(S) \sim \frac{4\pi}{3}r^3(\rho_{\rm eff}+3P_{\rm eff})
-\frac{\Lambda}{3}r^3.
\end{eqnarray}
Here note that the pressure term exists. Such term does not exist 
for Hayward's and Hawking's energies which have only 
the local energy density term
\cite{Small}\cite{Horowitz}. As the pressure can be source 
of gravity also in general relativity, our result is more reasonable 
than that for Hayward's and Hawking's energies. 

Next, we consider a small sphere in the Schwarzschild spacetime for an
example of the vacuum case. 
Here we adopt the isotropic coordinate for
Schwarzschild space-time:
\begin{eqnarray}
ds^2=-\Bigl( \frac{1-\frac{M}{2r'}}{1+\frac{M}{2r'}} \Bigr)^2 dt^2
+\Bigl(1+\frac{M}{2r'} \Bigr)^4 d{\bf x}'^2.
\end{eqnarray}
Let us consider a small sphere outside the black hole whose center is
located at ${\bf x'}= {\bf a}$.
We assume that the sphere has coordinate radius $r=r_0$ in the
transformed coordinate ${\bf x}={\bf x'}-{\bf a}$. In this coordinate, 
the metric becomes 
\begin{eqnarray}
ds^2 =-\Bigl( \frac{1-\frac{M}{2r'}}{1+\frac{M}{2r'}} \Bigr)^2 dt^2
+\Bigl(1+\frac{M}{2r'} \Bigr)^4 (dr^2 + r^2 d\Omega_2^2)
\end{eqnarray}
where $r'^2 = r^2 + a^2 + 2ar\cos\theta$ and ${\rm cos} \theta 
:={\bf a} \cdot {\bf r}/|{\bf a}||{\bf r}|$. Here we assume that the
initial velocity of the surface is zero, so the ``centrifugal force
component'' $\omega_a = n_c h_a^b \nabla_b u^c = 0$.
Thus the effective mass can be expressed as
\begin{eqnarray}
M_{\rm eff} &= &\Bigl(\frac{A}{4\pi}\Bigr)^{\frac12}
\int_S \frac{dS}{16\pi} (C_{abcd}h^{ac}h^{bd} - 2\omega_a\omega^a) \\
&= & \Bigl(\frac{A}{4\pi}\Bigr)^{\frac12}
\int_S \frac{dS}{16\pi} R_{abcd}h^{ac}h^{bd}.
\end{eqnarray}
Using the extended Gauss-Codacci relation, we have
\begin{equation}
R_{abcd}h^{ac}h^{bd} = {}^{(2)}R + \sum_{i=t,r}g^{ii}(K_{iA}^BK_{iB}^A
- (K_{iA}^A)^2)
\label{curvature}
\end{equation}
where $A, B$ run on $\theta, \phi$ and $K_{iAB}$ is the second
fundamental forms of the normal vectors $\partial_t$ and $\partial_r$.

By virtue of the Gauss theorem $\int_S {}^{(2)}R = 8\pi$, we can
arrive at the form
\begin{equation}
\int_S \frac{dS}{16\pi} R_{abcd}h^{ac}h^{bd} = \frac12 - \frac14
\int_{-1}^{1}d(\cos \theta) \biggl[1 -
\Bigl(1+\frac{M}{2r'}\Bigr)^{-1}\frac{M}{r'^3}r(r+a\cos \theta) \biggr]^2
\label{integral}
\end{equation}
without much effort.

This integral can be evaluated analytically by converting
the integral variable from $\cos \theta$ to $r'$. In the case $r=r_0<a$,
\begin{eqnarray}
\int_S \frac{dS}{16\pi} R_{abcd}h^{ac}h^{bd} &= &
\frac{1}{am}(a^2-r_0^2+m^2) + \frac{3}{4am^2r_0}(F(m)-F(0)) - \frac12 +
\frac{1}{2am}\frac{(a^2-r_0^2-m^2)^2}{(a+m)^2-r_0^2}
\label{exact} \\
&= & -\frac45 \Bigl(\frac{m}{a}\Bigr)^2
\Bigl(1+\frac{m}{a}\Bigr)^{-4} \Bigl(\frac{r_0}{a}\Bigr)^4\Bigl[
1+ O\Bigl[ \Bigl( \frac{r_0}{a} \Bigr)^2 \Bigr]\Bigr]
\label{leading}
\end{eqnarray}
where $m=M/2, F(m):=(a^2-r_0^2-m^2)(a^2-r_0^2+m^2)\log(a+r_0+m)/(a-r_0+m)$.

The area $A$ can be estimated by similar method and
\begin{eqnarray}
A &= &2\pi \int_{-1}^1 d(\cos \theta) \Bigl(1+\frac{M}{2r'}\Bigr)^4r^2
\label{integral2} \\
&= & 4\pi r_0^2 \Bigl( 1 + \frac{4m}{a} +
\frac{3m^2}{ar_0}\log\frac{a+r_0}{a-r_0} + \frac{4m^3}{a(a^2-r_0^2)} +
\frac{m^4}{(a^2-r_0^2)^2} \Bigr) \label{exact2} \\
&= & 4\pi r_0^2 \Bigl(1+\frac{m}{a}\Bigr)^4 \Bigl[1 + O\Bigl[
\Bigl( \frac{r_0}{a} \Bigr)^2 \Bigr] \Bigr].
\label{leading2}
\end{eqnarray}

Hence the effective mass is
\begin{equation}
M_{\rm eff} = -\frac15 \Bigl(\frac{M}{a}\Bigr)^2
\Bigl(1+\frac{M}{2a}\Bigr)^{-2} \Bigl(\frac{r_0}{a}\Bigr)^4r_0 
\Bigl[1 + O\Bigl[
\Bigl( \frac{r_0}{a} \Bigr)^2 \Bigr] \Bigr].\quad (r_0<a)
\end{equation}
The negativity means that the effect of the tidal force along the
direction of ${\bf a}$ which prolongs the sphere dominates over the
effect of the tidal force normal to ${\bf a}$ which squeezes the
sphere. That is, on the whole, a small sphere which does 
not enclose the central black hole must expand due to the tidal force 
at the first moment. When one considers a sufficient small sphere in 
vacuum spacetimes, the gravity is too week and then the small sphere 
cannot collapse gravitationally at the first moment. The negativity 
of the effective mass on the small sphere reflect certainly such kind 
of reasonable feature. 

On the other hand, on the whole, a large sphere which 
encloses black hole should
shrink at the first moment, so the effective mass is expected to positive.
Actually we can obtain the exact expression for $M_{\rm eff}$ for this 
case, too. The above integral (\ref{integral}) is valid for large
sphere $r_0>a$ as long as the sphere does not intersect with the
horizon and gives 
\begin{equation}
M_{\rm eff} = M -
\Bigl(1+\frac{2M}{r_0}\Bigr)^{-2}\frac{a^2M^2}{r_0^3} 
\Bigl[1 + O\Bigl[
\Bigl( \frac{a}{r_0} \Bigr)^2 \Bigr] \Bigr]
 \quad
(r_0>a), 
\end{equation}
which confirms the above expectation.



\section{Quasi-local energy based on the $(2+2)$ formalism\label{def}}
Since the construction of the effective mass comes from the dynamics
of the test 2-surfaces, one can expect that the similar form can be
obtained by another procedure in which 2-surfaces is basic tool. In
this section, we will show that our quasi-local energy is in fact
similar to the quasi-local energy derived from the total Hamiltonian
of the $(2+2)$ formalism under the special choice of the gauge. The
formalism is just the double null foliation of 2-surfaces and then
this is a good example for the demonstration.

\subsection{Brief review on the $(2+2)$ formalism\label{math}}

In this sub-section, we review the $(2+2)$ formalism and introduce the
Hamiltonian. Broadly, we follow Ref.\cite{Hay1}. We take two
commutable vector fields $u^a$ and $v^a$, and regard them as the
evolution vectors. One can take the evolutional direction to be null
in the neighbourhood of regular region without loss of generality. This
``double null foliation'' is assumed hereafter. We take the parameters
of $u^a$ and $v^a$ to be $\xi$ and $\eta$. Then the 2-surfaces
$\{S_{\xi, \eta}\}$, generated by Lie-propagating a fixed
two-dimensional spacelike surface $S$ along $u^a$ and $v^a$, serve as
a foliation of the spacetime in the neighbourhood of $S$. We take the
origins of the parameters $\xi$ and $\eta$ so that $S_{0,0}$ coincides
with $S$. Since we assumed the double null foliation, we restrict
ourselves to the cases such that $u^a$ and $v^a$ give null
three-surfaces. Define $h_{ab}$, the induced metric on $S_{\xi,\eta}$,
and $r_a \equiv h_{ab}u^b$, $s_a \equiv h_{ab}v^b$ and $m \equiv -
\log(-(u-r)_a (v-s)^a)$. It is easy to see that $u-r$ and $v-s$ are
the null normal vectors to the foliation. Thus, using these
quantities, the metric can be written as
%

\begin{displaymath}
g_{ab} = h_{ab} - e^m( (u-r)_a (v-s)_b + (v-s)_a (u-r)_b).
\end{displaymath}
%
The dynamical equations for the system are derived from the
variational principle. The Lagrangian $L$ for the $(2+2)$ formalism is
obtained by expressing the four dimensional Lagrangian $^{(4)}L$ in
terms of the quantities defined on $S_{\xi,\eta}$:
%
\begin{equation}
\int ^{(4)}L = \int d\xi d\eta \int_{S_{\xi,\eta}} L,
\label{action}
\end{equation}
%
where $^{(4)}L$ is the sum of the Einstein-Hilbert Lagrangian
${}^{(4)}\epsilon R/16\pi$ and the matter Lagrangian $^{(4)}L_m$ and
$^{(4)}\epsilon$ is four-dimensional volume form and $R$ is the Ricci
scalar of the spacetime metric.

It is easily verified that the dynamical equations obtained by
extremizing the action integral (\ref{action}) are written as the
Euler-Lagrange equations
%
\begin{equation}
{\pounds_u} \frac{\delta L}{\delta {\pounds_u q}} +
{\pounds_v} \frac{\delta L}{\delta {\pounds_v q}} -
\frac{\delta L}{\delta q} = 0,
\label{EulerLagrange}
\end{equation}
%
where $q$ denotes the dynamical degrees of freedom such as $h_{ab},
r^a, s^a, m$ and matter fields.

It is also easy to verify that the Euler-Lagrange equations
(\ref{EulerLagrange}) are equivalent to a Hamiltonian system. The
Hamiltonian ${\cal H}$ is given by
%
\begin{displaymath}
{\cal H} = p\pounds_u q + {\hat p}\pounds_v q - L,
\end{displaymath}
%
where $p$ and ${\hat p}$ are the canonical momenta defined by
%
\begin{displaymath}
p \equiv \frac{\delta L}{\delta \pounds_u q},\quad
{\hat p} \equiv \frac{\delta L}{\delta \pounds_v q}.
\end{displaymath}
%
The gravitational part of the Lagrangian which follows from the
Einstein-Hilbert Lagrangian with the cosmological constant, after
removing the total divergence, becomes
%
\begin{eqnarray*}
L_G &=& \mu\frac{e^{-m}}{16\pi} \biggl[ {\cal R} + \frac{e^m}{2}
\Bigl( {}^{\rm Tr}(\pounds_{u-r} h \pounds_{v-s} h) -
{}^{\rm Tr}\pounds_{u-r} h {}^{\rm Tr}\pounds_{v-s} h\\
& &{}+ {}^{\rm Tr}{\pounds_{u-r} h}{\pounds_{v-s} m} +
{}^{\rm Tr}{\pounds_{v-s} h}{\pounds_{u-r} m} \Bigr) +
\frac12 D_a m D^a m + 2\omega_a \omega^a - 2\Lambda \biggr],
\end{eqnarray*}
%
where $\mu$, ${\cal R}$ and $D_a$ are the area two-form, the scalar
curvature and the covariant derivative of the 2-surface
$S_{\xi,\eta}$, respectively. We also define the following quantities:
%
\begin{eqnarray}
{}^{\rm Tr}\pounds_u h &\equiv& h^{ab}\pounds_u h_{ab}\nonumber\\
{}^{\rm Tr}\pounds_v h &\equiv& h^{ab}\pounds_v h_{ab}\nonumber\\
{}^{\rm Tr}(\pounds_u h \pounds_v h) &\equiv&
h^{ab}h^{cd}\pounds_u h_{ac} \pounds_v h_{bd}\nonumber\\
\omega_a &\equiv& \frac{h_{ab}}{2}\frac{[u-r, v-s]^b}{(u-r)^c(v-s)_c}.
\label{omegadef}
\end{eqnarray}
%
We note that the ``centrifugal force component'' $\omega_a$,
introduced in the previous section, coincides with the twist
$\omega_a$ defined here for a particular choice of the foliation. See
Appendix A for proof.

\subsection{Quasi-local energies derived from $(2+2)$ formalism}

In the standard $(3+1)$ formalism, the total energy associated with an
asymptotically flat spacelike hypersurface is defined as the integral
of the Hamiltonian over that hypersurface\cite{RT}. The analogous
quantity for the $(2+2)$ formalism is the integral of the Hamiltonian
${\cal H}$ over the surface $S$. We would like to relate the integral
$\int_S {\cal H}$ to the quasi-local energy associated with $S$.

Since the integral $\int_S {\cal H}$ does not have the correct
dimension of energy and is dimensionless (in geometrical units $c =
G_{\rm N} = 1$), we multiply it with the area radius $r \equiv
(A/4\pi)^{1/2}$ where $A$ is the area of $S$. However, the quantity
$r\int_S {\cal H}$ still cannot be viewed as the quasi-local energy of
$S$ without restriction since its value is not a geometrically
invariant quantity. Rather, it depends on the choice of the foliation
surfaces around $S$. So we try to fix this ambiguity.

Of the geometrical quantities introduced in section \ref{math}, $r^a$,
$s^a$ and $m$ actually represents the coordinate freedom and can be
set to zero on any particular surface $S$. More precisely, for any
given spacelike 2-surface $S$, one can always find a double null
foliation around $S$ such that $r^a=s^a=0, m=0, \nabla_a m=0$ on $S$
(See Appendix B). Note that they cannot be set to zero throughout the
foliation. Under this gauge, for example, the gravitational part of the 
Hamiltonian
${\cal H}_G$, which is derived from the gravitational part of the
Lagrangian $L_G$, becomes simple:
%
\begin{eqnarray}
{\cal H}_G &=& \frac{\mu}{16\pi}\biggl[-{\cal R} - \frac12
\Bigl({}^{\rm Tr}{\pounds_u h}
{}^{\rm Tr}{\pounds_v h} - {}^{\rm Tr}({\pounds_u h}{\pounds_v
h})\Bigr) + 2\omega_a\omega^a + 2\Lambda\biggr]\\
&=& \frac{\mu}{16\pi}\Bigl(-C_{abcd}h^{ac}h^{bd} - R_{ab}h^{ab} +
\frac{R}{3} + 2\omega_a\omega^a + 2\Lambda \Bigr),
\end{eqnarray}
%
where $C_{abcd}$, $R_{ab}$ and $R$ are the four-dimensional Weyl
tensor, Ricci tensor and Ricci scalar, respectively\footnote{
There still
remains a gauge freedom even after imposing the above gauge conditions
on $S$. The gauge condition $r^a = s^a = 0$ only fixes the internal
coordinate of $S_{\xi, \eta}$. The condition $m=0,
\nabla_a m=0$ fixes the foliation around $S$ along the direction of
the null normals while the coordinate on the surface $S$ is not fixed. 
The remaining freedom is manifested on $S$ as a rescaling of the null
normals $u-r$ and $v-s$:
%
\begin{equation}
u-r \to (u-r)' = e^\alpha (u-r),\quad v-s \to (v-s)' = e^{-\alpha}
(v-s).
\label{gaugetransf}
\end{equation}
%
Note that this freedom is not expressed as such a simple rescaling on
the foliation surfaces other than $S$ because the surfaces do not
coincide between the different gauges in general.}.

Now having fixed some gauge, we define the quasi-local energy by the
total Hamiltonian as follows:
%
\[
E(S) \equiv - \Bigl( \frac{A}{4\pi} \Bigr)^{\frac12} \int_S {\cal H}.
\]
%
This is an analogue of the ADM energy, which is constructed from the
total Hamiltonian in $(3+1)$ formalism.  

Recently Hayward\cite{Hay2}
defined his quasi-local energy as \footnote{
Although Hayward did not state explicitly in \cite{Hay2}, the
expression (\ref{Hayward2}) still depends on the choice of the
foliation around $S$. Since the the null normals $n^+_a \equiv -
e^m(u-r)_a$ and $n^-_a \equiv -e^m(v-s)_a$ are the gradients $n^+ =
d\eta, n^- = d\xi$, the twist $\omega_a$ can be rewritten as
\[
\omega_a = \frac{h_a^b(n^{+c}\nabla_b n^-_c - n^{-c}\nabla_b n^+_c)}
{2n^{+d}n^-_d},
\]
which changes under the transformation (\ref{gaugetransf}) as
$\omega_a \to \omega_a - D_a \alpha$.}
%
\begin{eqnarray}
E_{\rm Hay}(S) &\equiv& - \Bigl( \frac{A}{4\pi} \Bigr)^{\frac12} \int_S
{\cal H}_G\Bigm|_{\Lambda = 0} \nonumber\\
&=& \Bigl( \frac{A}{4\pi} \Bigr)^{\frac12} \int_S \frac{\mu}{16\pi}
\Bigl( C_{abcd}h^{ac}h^{bd} + R_{ab}h^{ab} - \frac{R}{3} - 2\omega_a
\omega^a \Bigr).
\label{Hayward2}
\end{eqnarray}
%
The difference with $E(S)$ is that Hayward has used only the gravitational 
part of the Hamiltonian without the cosmological constant, not the total one. 
When there is no matter field and the cosmological constant vanishes,
$E(S)$ reduce to $E_{\rm Hay}(S)$ if we impose Ricci flat condition.
As the  calculation is rather complicated for the cases with matters in the 
double null formalism, we concentrate on the vacuum case here. That
is, if there is only cosmological constant $\Lambda$ and no matter
fields exist, the quasi-local energy $E(S)$ is
%
\begin{equation}
E(S) = \Bigl( \frac{A}{4\pi} \Bigr)^{\frac12}
\int_S \frac{\mu}{16\pi}
\Bigl(C_{abcd}h^{ac}h^{bd} - 2\omega_a \omega^a - \frac43 \Lambda\Bigr).
\label{common}
\end{equation}
%
If one choose another gauge fixing, one will obtain an another form. 
However, the fact that we could obtain the same form with the effective 
mass defined in the section 2 is important. As we guessed, one can see 
that the effective gravitational
mass derived from the physical argument of freely falling 2-surface
really has a relation with the Hamiltonian energy derived from the
$(2+2)$ formalism.

The similarity of the expression between $M_{\rm eff}(S)$ and $E(S)$
should be remarked and we have the interpretation such that one can give 
support $M_{\rm eff}(S)$ from the theoretical point of view.


\section{Summary\label{conc}}

We have defined the quasi-local energy from the concept of the
effective gravitational mass for freely falling 2-surface $S$. Its
expression is given by 
%
\begin{equation}
M_{\rm eff}(S) =
\Bigl( \frac{A}{4\pi} \Bigr)^{\frac12}
\int_S \mu \biggl[
\frac{1}{16\pi} \Bigl( C_{abcd}h^{ac}h^{bd} - 2\omega_a \omega^a -
\frac43 \Lambda \Bigr)
+ T_{ab}t^a t^b - \frac12 T_{ab}h^{ab} + \frac23 T_a^a \biggr].
\end{equation}
%
It has
the advantage that the gravitational mass is related directly with the
motion of a body under free fall, so is well adapted to the intuitive
physical expectation. It is not obscured by mathematical complication
which sometimes covers over the quasi-local energies proposed so far. 
We have also found that it reduces to
the ADM energy and Bondi-Sachs energy at the infinity in the
asymptotically flat spacetimes.
In spherically symmetric spacetimes, it gives the appropriate
gravitational mass for radially infalling test particles. We also
found the similarity of the effective mass with the quasi-local energy
derived from the total Hamiltonian of the $(2+2)$ formalism in the
vacuum cases.

Furthermore, we evaluated the effective mass for small spheres. In the 
non-vacuum case, we obtain the leading term 
%
\begin{equation}
M_{\rm eff}(S) \sim \frac{1}{3}R_{ab}t^at^b
\end{equation}
%
and, in vacuum space-times without $\Lambda$ term, we 
observe the effective mass for small sphere outside the black hole is
negative in the Schwarzschild spacetime.
We discussed that the negativity is reasonable from the view point of 
the tidal force. Hence, for our effective energy, the negativity is 
no problem in spite of Bergqvist's claim. Rather one should prove that 
the effective mass must have the negative lower bound.

 From the construction we expect that the effective mass is useful to
investigate the dynamics of the space-time. The application will be
considered in the future work.

\section*{Acknowledgement}
We would like to thank Sean A. Hayward for his important suggestion
and discussion. We would also like to thank Katsuhiko Sato, Yasushi
Suto, Takahiro T. Nakamura and Gen Uchida for useful comments and
discussions.

\appendix
\section{Equivalence of two definitions of $\omega_a$}
In this appendix we show that two definitions of $\omega_a$, namely,
$n_ch_a^b\nabla_bt^c$ and eq.(\ref{omegadef}), are equivalent for a
particular choice of gauge.

Consider a 1-parameter family of 2-surfaces generated by
Lie-propagating the initial 2-surface $S$ along the free-fall vector
$t^a$. We would like to construct a double null foliation around $S$
some surfaces of which coincide with the members of this 1-parameter
family.

If such foliation exists, each 2-surface of the 1-parameter family is
a cross section of two null hypersurfaces generated by null geodesics
normal to 2-surfaces on $\pounds_u S$ and $\pounds_v S$. Conversely, 
if we generate two null hypersurfaces by null geodesics normal to a
member of the 1-parameter family and repeat this procedure for each
member of the family, we have the desired double null foliation around
$S$.

Choose two future directed null vector fields $n_\pm^a$ on $\pounds_t
S$ such that $n_+^an_{-a} = -1, (n_+^a + n_-^a)/\sqrt{2} = t^a$. Then
we have $n^a = (n_+^a - n_-^a)/\sqrt{2}$. Extend $n_+^a$ along its
direction by parallel transport: $n_+^a\nabla_an_+^b = 0$. Similarly,
extend $n_-^a$ in the same way: $n_-^a\nabla_an_-^b = 0$. Demanding
the normalization condition $n_+^an_{-a} = -1$, we have normalized
null normal fields to the double null foliation around $S$. Define
$t^a \equiv (n_+^a + n_-^a)/\sqrt{2}$ and $n^a \equiv (n_+^a -
n_-^a)/\sqrt{2}$ on the entire foliation.

Now we examine the $\omega_a$ for this foliation. Since $u-r$ and
$v-s$ in eq.(\ref{omegadef}) are the null normals to the foliation and 
proportional to $t^a - n^a$ and $t^a + n^a$, respectively, it is easy
to see that
\begin{eqnarray*}
\omega_a &=& \frac{h_{ab}}{2} \frac{[t-n, t+n]}{(t-n)^c(t+n)_c}\\
	&=& -\frac{h_{ab}}{2}(t^c\nabla_cn^b - n^c\nabla_ct^b)\\
	&=& n_ch_a^b\nabla_bt^c + h_a^bn^c\nabla_{[c}t_{b]}.
\end{eqnarray*}

Thus our aim is to show $X^an^b\nabla_{[b}t_{a]}=0$ on $S$ for an
arbitrary vector field $X^a$ tangent to $S$. Now we have
\begin{eqnarray}
2X^an^b\nabla_{[b}t_{a]} &=& t_a[X, n]^a\nonumber\\
	&=& (n_+^a + n_-^a)[X, n_+ - n_-]^a.
\label{expressiontobevanished}
\end{eqnarray}

Since $[X, n_+]^a$ is tangent to the null hypersurface $\pounds_v S$,
it is orthogonal to its normal vector $n_+^a$. A similar relation
holds for $n_-^a$, too:
\begin{equation}
n_{+a}[X, n_+]^a = n_{-a}[X, n_-]^a = 0.
\label{bothzero}
\end{equation}

On $\pounds_t S$, we also have
\begin{displaymath}
[t, X]^a = [n_+ + n_- , X]^a \parallel S.
\end{displaymath}
So, contracting $n_{+a}$ and $n_{-a}$ with this expression and using
eq.(\ref{bothzero}), we have
\begin{equation}
n_{+a}[X, n_-]^a = 0, n_{-a}[X, n_+]^a = 0.
\label{bothzero2}
\end{equation}

Eqs.(\ref{bothzero},\ref{bothzero2}) tell us the expression
(\ref{expressiontobevanished}) vanishes. This completes the proof.

\section{Achieving the gauge $r^a=s^a=0, m=0, \nabla_a m=0$ on $S$}
Here we show the existence of the double null foliation satisfying the
gauge condition $r^a=s^a=0, m=0, \nabla_a m=0$ on $S$, which was
briefly stated in \cite{Hay2}. First we show that there always exists
a double null foliation such that $r^a=s^a=0$ on $S$. Introduce
internal coordinates $(\vartheta,
\varphi)$ on $S_{\xi,\eta}$ so that the evolution vectors are
expressed as partial derivatives $u^a = (\partial_\xi)^a, v^a =
(\partial_\eta)^a$. They lie on the intersection of the 2-surface
$\vartheta, \varphi = \mbox{const}$ and the null hypersurfaces $\eta =
\mbox{const}$, $\xi = \mbox{const}$, respectively. Thus to show the
existence of the foliation with $r^a|_S=s^a|_S=0$, it suffices to show
the existence of a coordinate chart $(\vartheta, \varphi)$ such that
each 2-surface $\vartheta, \varphi = \mbox{const}$ is normal to $S$,
which is obvious.

Modifying this foliation, it is possible to achieve $m=0, \nabla_a
m=0$ on $S$. (Intuitively, this is obvious since $m$ represents the
`density' of the foliation surfaces.) In general, any two foliations
are related by a coordinate transformation $\xi \to \xi' = \xi f, \eta
\to \eta' = \eta g$ where $f, g$ are some smooth functions. The reason
why $\xi$ and $\eta$ can be factored out in $\xi'$ and $\eta'$ is that
the null hypersurfaces $\xi = 0$ and $\eta = 0$ are uniquely determined
from $S$ and do not depend on the choice of the foliation. Since the
inverse of the spacetime metric is expressed in the original
coordinate as
\begin{displaymath}
g^{ab} = -e^m ((u-r)^a (v-s)^b + (v-s)^a (u-r)^b) +
h^{ij}(\partial_i)^a (\partial_j)^b
\end{displaymath}
where the indexes $i,j$ run on $\{\vartheta$, $\varphi\}$,
\begin{eqnarray*}
e^{m'}|_S &= -g^{-1}(d\xi', d\eta') = fg e^m|_S\\
e^{m'}d m'|_S &= e^m (fg d m|_S + d(fg)|_S +
		f\partial_\eta g d\eta|_S + g\partial_\xi f d\xi|_S ).
\end{eqnarray*}
The condition $m'|_S = 0$ is thus equivalent to $fg|_S = e^{-m}$. 
With this condition satisfied, the condition $\nabla_a m'|_S = 0$ is
equivalent to $\partial_\xi m'|_S = \partial_\eta m'|_S = 0$ which is
further equivalent to $e^{-m} \partial_\xi m|_S + f\partial_\xi g|_S
+ 2g\partial_\xi f|_S = 0$ and $e^{-m} \partial_\eta m|_S +
g\partial_\eta f|_S + 2f\partial_\eta g|_S = 0$. These conditions only
specify the behaviour of $f,g$ and their first order derivatives on
$S$ and it is not difficult to see that they are compatible with the
double null conditions $g^{-1}(d\xi', d\xi') = g^{-1}(d\eta',
d\eta') = 0$. This establishes the existence of the desired foliation.

\end{document}